\documentclass[twocolumn,prd,floatfix]{revtex4}
\usepackage{graphicx}

\newcommand{\be}{\begin{equation}}
\newcommand{\e}{\end{equation}}
\newcommand{\bear}{\begin{eqnarray}}
\newcommand{\ear}{\end{eqnarray}}
\newcommand{\nline}{\nonumber \\}
\newcommand{\f}{\frac}
\newcommand{\de}{{\rm d}}
\newcommand{\del}{\partial}

\newcommand{\la}{\langle}
\newcommand{\ra}{\rangle}

\begin{document}


\title{Analytical Models of the Intergalactic Medium and Reionization}

\author{Tirthankar Roy Choudhury}

\email{Email: tirth@hri.res.in}  

\affiliation{Harish-Chandra Research Institute, Chhatnag Road,\\
Jhusi, Allahabad 211 019, India.}

\begin{abstract}
Reionization is 
a process whereby hydrogen (and helium) in the Universe is ionized by the
radiation from first luminous sources.
Theoretically, 
the importance of the reionization lies in its close coupling with the 
formation of first cosmic structures and hence there is considerable
effort in modelling the process. We 
give a pedagogic overview of different analytical approaches
used for modelling reionization. We also discuss 
different observations related to reionization and show how
to use them for constraining the reionization
history.
\end{abstract}

\maketitle



\section{Introduction}
\label{intro}

Current models of cosmology \cite{dkn++09} 
indicate that about three-fourth of the 
energy density in our present Universe is constituted by ``dark energy'' which 
is responsible for the current acceleration of the cosmic expansion. 
The next dominant component
is the ``dark matter'' which constitutes about 23 per cent of the density. 
This form of matter is collisionless and interact only gravitationally.
The baryons constitute only 2 per cent of the total mass.
The two most abundant elements among the baryons are hydrogen and helium.

Study of reionization mostly concerns with the ionization and thermal
history of the baryons (hydrogen and helium) in our Universe
\cite{lb01,bl01,cf06a,fck06}.
Within the framework of hot Big Bang model, 
hydrogen formed for the first time  
when the age of the Universe was about $3 \times 10^5$ years, its size being
one-thousandth of the present (corresponding to a scale factor 
$a \approx 0.001$
and a redshift $z = 1/a - 1 \approx 1000$). Around this
time, the temperature of the radiation 
became low enough $\approx 3 \times 10^3$ K
that the photons were not able to ionize the proton-electron
pair through collisions and hence formation of hydrogen 
(and some amount of helium) 
could take place \cite{peebles93}. 
The epoch at which the protons and the electrons 
combined for the first time to form hydrogen 
atoms is known as the recombination epoch and is well-probed
by the Cosmic Microwave Background Radiation (CMBR).

Right after the recombination epoch, the 
Universe entered a phase called the ``dark ages'' where no significant 
radiation sources existed. The hydrogen remained largely neutral
at this phase. The small inhomogeneities in the dark matter density field 
which were present during the recombination epoch started growing 
via gravitational instability giving rise to highly
nonlinear structures like the collapsed haloes \cite{cr86}. 
It should, however, be kept in mind that most of the baryons at high redshifts
do not reside within these haloes, they rather 
reside as diffuse gas within the intergalactic space which is
known as the intergalactic medium (IGM) \cite{peacock99,padmanabhan02b}.

The collapsed haloes form potential wells whose depth depend on their mass and 
the baryons (i.e, hydrogen) then ``fall'' in these wells.
If the mass 
of the halo is high enough (i.e., the potential well is deep enough), 
the gas will 
be able to dissipate its energy, cool via atomic or molecular
transitions and fragment within the halo.
This produces conditions appropriate for condensation
of gas and forming stars and galaxies. Once these
luminous objects form, the era of dark ages can be thought of 
being over.
 
The first 
population of luminous stars and galaxies 
can generate 
ultraviolet (UV) radiation through nuclear reactions. 
In addition to the galaxies, perhaps an early 
population of accreting black holes (quasars) also generated
some amount UV radiation. 
The UV radiation contains photons
with energies $> 13.6$ eV which are then able to ionize  hydrogen 
atoms in the surrounding medium, a process known as ``reionization''. 
Reionization is thus the second major change in the ionization state of 
hydrogen (and helium) in the Universe (the first being the recombination).

As per our current understanding \cite{gnedin00,wl03,cf06b}, 
reionization started around the 
time when first structures formed, which is currently believed 
to be around $z \approx 20-30$. In the simplest picture, each source 
first produced an ionized region around it; these
regions then overlapped and percolated into the IGM. This 
era is usually called the ``pre-overlap'' phase. The
process of overlapping 
seemed to be completed around $z \approx 6-8$ at which point
the neutral hydrogen fraction fell to values lower than
$10^{-4}$. Following that a never-ending ``post-reionization'' (or
``post-overlap'')
phase started which implies that the Universe is largely ionized 
at present epoch. Reionization by UV radiation is also accompanied
by heating: electron which are released by photoionization will 
deposit an extra energy 
equivalent to $h_p \nu - 13.6~{\rm eV}$ to the IGM, where $\nu$ is the
frequency of the ionizing photon and $h_p$ is the Planck constant.
This reheating of the IGM 
can expel the gas and/or suppress cooling in the low mass haloes 
-- thus, there is 
a considerable reduction in the  
cosmic star formation right after reionization. In addition, 
the nuclear reactions within the stellar sources potentially
alter the chemical composition of the medium if the star dies via 
energetic explosion (supernova). This can change the nature of star
formation at later stages.

The process of reionization is 
of immense importance in the study of structure formation since, on 
one hand,  
it is a direct consequence of the 
formation of first structures and luminous sources while, on the other, 
it affects subsequent structure formation. Observationally, the reionization
era represents a phase of the Universe which is yet to be probed; 
the earlier phases ($z \approx 1000$) are probed
by the CMBR while the post-reionization phase ($z < 6$) is probed
by various observations based on galaxies, clusters, quasars and other sources. 
In addition to the importance outlined above, the study of dark ages 
and cosmic reionization has acquired increasing 
significance over the last few years because of the availability of 
good quality data in different areas.

In this article, we will mainly concentrate on the current status
of various analytical and semi-analytical approaches 
which go into modelling reionization. The main aim would be to
systematically discuss the set of equations which are crucial
in understanding the process highlighting the major
physical processes and assumptions. We shall also
highlight the relevant observational probes at appropriate places.
In Section \ref{chap2}, we shall give a pedagogic 
introduction to the basic theoretical formalism
for studying reionization and IGM in different phases of evolution.
Section \ref{chap3} would be devoted to discussing detailed modelling of 
reionization using the formalism developed. We shall illustrate
on how to constrain the models by comparing with a wide variety of 
available data sets.
In Section \ref{chap4}, we shall briefly discuss the 
current numerical simulations and observations related to reionization. We shall
also highlight what to expect in this field in near future.


\section{Theoretical Formalism}
\label{chap2}

In this section, we discuss the basic theoretical formalism 
required for modelling reionization of the IGM. 
The main aim here would be to highlight
the physical processes which are crucial in understanding reionization and
comparing with observations. In what follows, we shall assume that
the IGM consists only of hydrogen and neglect the presence of helium. It is
straightforward to include helium into the formalism.

Essentially, in presence of a ionizing radiation, 
the evolution of the mean neutral hydrogen density $n_{\rm HI}$ \footnote{In 
astrophysical notation, HI stands for neutral hydrogen while 
HII denotes ionized hydrogen (proton).} is given by
\be
\dot{n}_{\rm HI} = -3 H(t) n_{\rm HI} 
- \Gamma_{\rm HI} n_{\rm HI} + {\cal C} \alpha(T) n_{\rm HII} n_e
\label{eq:dnhidt_global}
\e
where overdots denote the total time derivative $\de/\de t$, 
$H \equiv \dot{a}/a$ is the Hubble parameter, 
$\Gamma_{\rm HI}$ is the photoionization rate per hydrogen atom, 
$\alpha(T)$ is the recombination rate coefficient and 
$n_e$ represents the mean electron density. 
The first term in the right hand side of equation (\ref{eq:dnhidt_global})
corresponds to the
dilution in the density because of cosmic expansion, the second term
corresponds to photoionization by the ionizing flux 
and the third term corresponds to recombination of protons and
free electrons into neutral hydrogen. 
The quantity ${\cal C}$ is called the clumping factor and is defined as
\be
{\cal C} \equiv \f{\la n_{\rm HII} n_e \ra}{\la n_{\rm HII}\ra \la n_e \ra} = 
\f{\la n_H^2 \ra}{\la n_H\ra^2}
\e
where the last equality holds for the case when the IGM contains only
hydrogen (i.e., no helium) and is highly
ionized, i.e, $n_e = n_{\rm HII} \approx n_H$. The clumping factor takes
into account the fact that the recombination rate in an inhomogeneous
(clumpy) IGM is higher than a medium of uniform density. 

The ionization equation is usually supplemented
by the evolution of the IGM temperature $T$, which is given by
\be
\dot{E}_{\rm kin} = -2 H(t) E_{\rm kin} + \Lambda 
\label{eq:dEkindt_global}
\e
where $E_{\rm kin} = 3 k_B T n_H$ is the kinetic energy of
the gas and $\Lambda$ is the net heating rate including 
all possible heating and cooling processes. The first term on the 
right hand side takes into account the adiabatic cooling of the gas
because of cosmic expansion.

\subsection{Cosmological radiation transfer}

The equation of radiation transfer, 
which describes propagation of radiation flux through
a medium, is written as an evolution equation for the 
specific intensity of radiation 
$I_{\nu} \equiv I(t,{\bf x},\nu,{\bf \hat{n}})$ which has dimensions of
the energy per unit time per unit area per unit solid angle 
per frequency range. 
It is a function of time and space coordinates $(t, {\bf x})$, 
the frequency of radiation $\nu$ and the direction of propagation 
${\bf \hat{n}}$.
The radiation transfer equation in a cosmological scenario has the form 
\cite{anm99}
\bear
\f{\del I_{\nu}}{\del t} + \f{c}{a(t)} {\bf \hat{n} \cdot \nabla_x} I_{\nu}
-H(t) \nu \f{\del I_{\nu}}{\del \nu} 
+ 3 H(t) I_{\nu}
\nline
= -c \kappa_{\nu} I_{\nu} + \f{c}{4 \pi} \epsilon_{\nu},
\label{eq:radtrans_local}
\ear
where $\kappa_{\nu}$ is the absorption coefficient and $\epsilon_{\nu}$ is the
emissivity. The above equation is essentially the Boltzmann equation
for photons with $I_{\nu}$ being directly proportional to the 
phase space distribution function \cite{padmanabhan02b}. 
The terms on the left hand side of 
equation (\ref{eq:radtrans_local}) add up to the total time
derivative of $I_{\nu}$; in particular, the third term corresponds 
to dilution of the intensity and 
the fourth term accounts for shift of frequency $\nu \propto a^{-1}$
because of cosmic expansion. 
The effect of scattering (which is 
much rarer
than absorption in the IGM) can, in principle, 
be included in the $\kappa_{\nu}$
term if required. 
If the medium contains absorbers with number density 
$n_{\rm abs}$ each having a cross-section $\sigma_{\nu}$, the absorption
coefficient is given by $\kappa_{\nu} = n_{\rm abs} \sigma_{\nu}$. The mean
free path of photons in the medium is given by $\lambda_{\nu}(t) \equiv 
\kappa^{-1}_{\nu}(t)$.

We define the mean specific intensity by averaging $I_{\nu}$ 
over a large volume and over all directions
\be
J_{\nu}(t)
\equiv \int_V \f{\de^3 x}{V} \int \f{\de \Omega}{4 \pi}  
I_{\nu}(t,{\bf x},{\bf \hat{n}})
\e
Then the spatially and angular-averaged radiation transfer equation becomes
\cite{peebles93}
\be
\dot{J}_{\nu} \equiv
\f{\del J_{\nu} }{\del t} 
-H(t) \nu \f{\del J_{\nu} }{\del \nu} 
= - 3 H(t) J_{\nu} 
-c \kappa_{\nu} J_{\nu}  + \f{c}{4 \pi} \epsilon_{\nu}
\e
where the coefficients $\kappa_{\nu}$  and $\epsilon_{\nu}$ 
are now assumed to be averaged over the large volume. The quantity 
$J_{\nu}$ is essentially the energy per unit time per unit area
per frequency interval per solid angle.

The integral solution of the above equation along a line of sight
can be written as \cite{hm96}
\be
J_{\nu}(t) = \f{c}{4 \pi} \int_0^t \de t' 
\epsilon_{\nu'}(t')
\left[\f{a^3(t')}{a^3(t)}\right] 
{\rm e}^{-\tau(t,t';\nu)}.
\label{eq:jnu_intsol_tau}
\e
where $\nu' = \nu a(t)/a(t')$,  $\nu'' = \nu a(t)/a(t'')$
and 
\be
\tau(t,t';\nu) \equiv c \int_{t'}^t \de t''
\kappa_{\nu''}(t'')
= c \int_{t'}^t \f{\de t''}{\lambda_{\nu''}(t'')}
\e
is the optical depth along the line of sight from $t'$ to $t > t'$
Clearly, the intensity at a given epoch is proportional 
to the integrated emissivity with an exponential attenuation 
due to absorption in the medium.
The intensity attenuates by $1/{\rm e}$ when the radiation travels a distance
equal to the mean free path.

The absorption is ``local'' when the mean free path of photons 
is much smaller than the horizon size of the Universe, i.e., 
$\lambda_{\nu}(t) \ll c/H(t)$. 
In addition, if we also assume that the emissivity $\epsilon_{\nu}$
does not evolve
significantly over the small time interval $\lambda/c$, then the specific
intensity is related to the emissivity through a simple form
\cite{mhr99,mhr00}
\be
J_{\nu}(t) \approx \f{\epsilon_{\nu}(t) \lambda_{\nu}(t)}{4 \pi}
= \f{\epsilon_{\nu}(t)}{4 \pi \kappa_{\nu}(t)}
\label{eq:jnu_mfp}
\e
Note that in the case of local absorption, $\dot{J}, H J  \ll c \epsilon$.
In this approximation, the background intensity 
depends only on the instantaneous value of the 
emissivity (and not its history) because all the photons are
absorbed shortly after being emitted (unless the 
sources evolve synchronously over a timescale much
shorter than the Hubble time). 
We shall discuss later in Section \ref{chap2:postreion} 
that this is a useful approximation
for the IGM for redshifts $z \gtrsim 3$.

\subsection{Post-reionization epoch}
\label{chap2:postreion}

Let is first study the radiation transfer in the 
post-reionization epoch. Compared to the 
pre-overlap era, this epoch is much easier to study because 
the IGM can be treated as
a highly ionized single-phase medium (whereas during the pre-overlap era,
one is looking into two distinct phases -- ionized and neutral).
The optical depth can be written as
\be
\tau(z,z';\nu) = c \int_{z}^{z'} \f{\de z''}{(1+z'') H(z'')} 
n_{\rm HI}(z'') ~ \sigma_{\rm abs}(\nu'')
\e
where $\sigma_{\rm abs}(\nu)$ is the total absorption cross section 
of neutral hydrogen and we have changed the time coordinate to the redshift 
$z$. 
Various processes can, in principle, contribute
to $\sigma_{\rm abs}(\nu)$, most dominant being the resonant 
Lyman series absorption
corresponding to excitation of hydrogen atoms from the ground
state to higher ones (1s $\to n$p) and the
continuum absorption of photons above the ionization threshold via
photoionization process.
Let us treat each of them separately in the following:

\subsubsection{Resonant Lyman series absorption}
\label{chap2:resonant}

The Lyman series absorption
arises from the electronic excitation of neutral hydrogen atoms 
from the 1s ground
state to higher ones. The most dominant of these are the
Ly$\alpha$ (1s $\to$ 2p, rest wavelength $\lambda_{\alpha} \approx 1216 \AA\ $) 
and Ly$\beta$ (1s $\to$ 3p, $\lambda_{\beta} \approx 1206 \AA\ $) 
transitions, and hence they are the most relevant ones 
as far as observations are concerned. For simplicity,
we shall present results for the Ly$\alpha$ absorption only, the others can be
calculated in identical manner. The Ly$\alpha$ absorption 
cross section is given by
\be
\sigma_{\rm abs}(\nu) = \sigma_{\alpha}~
V\left(\f{\nu}{\nu_{\alpha}} - 1\right)
\label{eq:sigma_alpha}
\e
where $\nu_{\alpha} = c/\lambda_{\alpha}$ is the resonant frequency of transition,
$\sigma_{\alpha} = 4.45 \times 10^{-18} {\rm cm}^2$ is the 
cross section at $\nu_{\rm alpha}$ 
and  $V$ is a function which determines the profile
of the absorption line. It is called the Voigt profile function and is 
a convolution of the Lorentzian shape for the natural 
broadening and the Gaussian shape for the thermal broadening.
For the purpose of this article, it
is sufficient to note that $V$ is a sharply peaked function
about $\nu/\nu_{\alpha} = 1$; for most our discussion, we shall take it to be 
a Dirac-delta function $V(\nu/\nu_{\alpha} - 1) = \delta_D(\nu/\nu_{\alpha} - 1)$.

The optical depth between the redshifts $z$ and $z'$ is then given by
\bear
\tau(z,z';\nu) = \tau(z_{\alpha}) =  
\sigma_{\alpha} \f{c}{H(z_{\alpha})}
n_{\rm HI}(z_{\alpha});
\nline
1+z_{\alpha} = \f{\nu_{\alpha}}{\nu} (1+z)
\ear
If we put this into equation (\ref{eq:jnu_intsol_tau}), we see that the
Ly$\alpha$ absorption at a redshift $z_{\alpha}$  reduces the specific
intensity observed at $z$ at a frequency $\nu_{\alpha} (1+z)/(1+z_{\alpha})$
by a factor $e^{-\tau(z_{\alpha})}$. The value of $\tau(z_{\alpha})$ 
along a given line of sight would depend upon the distribution of 
$n_{\rm HI}(z_{\alpha})$. However, we would mostly be interested in the mean
value of specific intensity averaged over a number of lines of sight. The
corresponding reduction can be described by a line-of-sight-averaged
optical depth
\be
{\rm e}^{-\tau_{\rm eff}(z_{\alpha})} \equiv 
\la {\rm e}^{-\tau(z_{\alpha})} \ra_{\rm LOS}
\e
where $\la~\ra_{\rm LOS}$ denotes averaging over lines of sight. 
The quantity $\tau_{\rm eff}$ is usually known as the 
``effective optical depth''. 

Theoretically, the value of $\tau_{\rm eff}$ can be calculated if we 
know the distribution of optical depth $P(\tau)$ [which can be calculated
from the neutral hydrogen distribution $P(n_{\rm HI})$]:
\be
{\rm e}^{-\tau_{\rm eff}(z)} =
\int_0^{\infty} \de \tau ~ P(\tau; z) ~ {\rm e}^{-\tau}
\e
Of course, one requires detailed understanding of the evolution of the
baryonic density field to model the distribution $P(n_{\rm HI})$. This
has been addressed by a series of analytical studies 
\cite{bbc92,bi93,gh96,bd97,hgz97,gh98,mhr00,cps01,csp01}
and numerical
simulations \cite{cmor94,zan95,hkwm96,mcor96,vmmth02,msc++05,bhvs05,msb++06,bvkhc08}, which we shall avoid discussing here. 
However, we can still make some inference assuming the distribution
is uniform, i.e., $\tau_{\rm eff}(z) = c \sigma_{\alpha} 
n_{\rm HI}(z)/H(z)$. 
If we define the neutral hydrogen fraction to be
$x_{\rm HI} \equiv n_{\rm HI}/n_H$, then we  can calculate
$\tau_{\rm eff} \propto x_{\rm HI}$ given a set of cosmological
parameters (which would uniquely determine $H(z)$ and $n_H$).

\begin{figure}
\rotatebox{270}{\includegraphics[width=6cm]{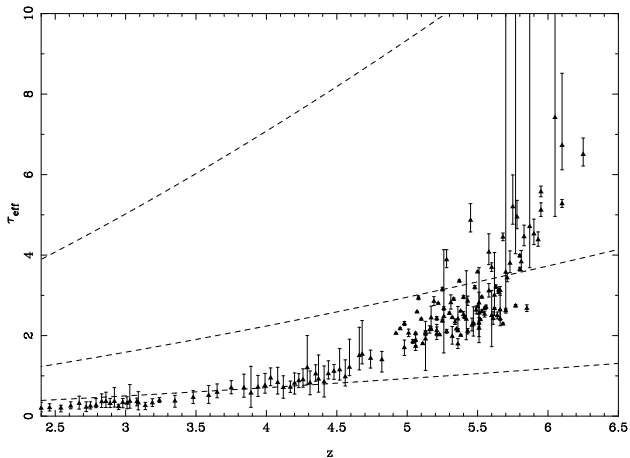}}
\caption{The effective optical depth of Ly$\alpha$ absorption as function
of redshift $z$. The points with errorbars represent the observational data.
The dashed curves, from top to bottom, represent the predictions for 
a uniform IGM with neutral hydrogen fraction $x_{\rm HI} = (3, 1, 0.3) \times 10^{-5}$, respectively.}
\label{fig:tauGP}
\end{figure}

Observationally, $\tau_{\rm eff}$ can be determined by looking at 
the spectra of bright sources like quasars at high redshifts. These spectra
show a series of absorption features at frequencies 
larger than the Ly$\alpha$ frequency in the quasar rest frame.
Since one has a good knowledge of the unabsorbed quasar
spectra (from looking at nearby quasars and also having
some understanding about the physical processes), one can
calculate the amount of absorption happening
because of the intervening IGM between the quasar and the observer; this
absorption, averaged over numerous lines of sight, 
is essentially the quantity ${\rm e}^{-\tau_{\rm eff}}$. This has been
done to quite high redshifts $z \sim 6.5$ and the values
of $\tau_{\rm eff}$ observed are shown as points with errorbars 
\cite{songaila04,fsb++06} 
in Figure \ref{fig:tauGP}.

To understand what these values imply, we have plotted with dashed lines
the calculated value of $\tau_{\rm eff}$ for a uniform IGM assuming three
values of $x_{\rm HI} = (3, 1, 0.3) \times 10^{-5}$ from top to bottom. 
This immediately tells us that the fraction of neutral hydrogen
has to be $\sim 10^{-5}$ in order to reproduce the observed values of
$\tau_{\rm eff}$. In fact, had $x_{\rm HI}$ been slightly (say $\sim 10^{-4}$) 
higher, one would have obtained $\tau_{\rm eff}$ much higher 
than unity ($\sim 10-100$) and
hence the flux from the quasar would be completely absorbed. If that
were the case, it would
show up as a absorption ``trough'' at frequencies larger 
than the rest frame Ly$\alpha$ frequency. 
In reality, a considerable amount of transmitted flux is
found at these frequencies alongwith a series
of absorption features arising from the Ly$\alpha$ transition of
residual neutral hydrogen. These
absorption signatures are known as the ``Ly$\alpha$'' forest and are
powerful probes of the neutral hydrogen distribution in the IGM at $z<6$
\cite{rauch98}.

The absence of  a absorption
trough is a direct proof of the fact that hydrogen is 
completely reionized in the diffuse IGM 
at redshifts $z \lesssim 6$. This is known as the 
Gunn-Peterson effect \cite{gp65}. Note that the actual inferred 
value of $x_{\rm HI}$ might be
slightly different if one models with an appropriate density distribution,
however, the basic conclusion remains unchanged.

We can also see from Figure \ref{fig:tauGP} that for quasars at 
redshifts $z \gtrsim 6$, the observed value of 
$\tau_{\rm eff} \gtrsim 5$; this would imply an attenuation $\gtrsim 0.01$
and hence one actually observes absorption troughs as predicted
by Gunn-Peterson effect \cite{fsb++06,wdr++09}. Unfortunately, 
finding such troughs does not
necessarily imply that the IGM is highly neutral as even a 
$x_{\rm HI} \sim 10^{-4}$ could be sufficient to absorb all the flux. However,
one can use much detailed modelling to improve the constraint, which we shall
discuss later in Section \ref{chap3_cfmodel}.

The values of $\tau_{\rm eff} \lesssim 1$ at $z < 4$ means that the diffuse IGM 
is highly transparent (also called optically thin) 
to Ly$\alpha$ photons. Only about $\sim 10\%$ of the Ly$\alpha$ 
photons are absorbed, mostly within the high density regions.
These high density systems are often modelled as a set
of discrete absorbers of some size. If we consider an absorber 
having neutral hydrogen density $n_{\rm HI}$ and a size $L \ll c/H(z)$ along
the line of sight at a redshift $z_{\rm abs}$, the optical depth is given by
\be
\tau(z,z';\nu) = N_{\rm HI} ~ \sigma_{\rm abs}(\nu_{\rm abs})
\e
where $\nu_{\rm abs} = \nu (1+z_{\rm abs})/(1+z)$ and 
$N_{\rm HI} \equiv n_{\rm HI} ~ L$ is the column density of neutral 
hydrogen within the absorber. 
Hence, each absorber reduces the specific intesnity by a factor
${\rm e}^{-N_{\rm HI} \sigma_{\rm abs}(\nu_{\rm abs})}$. 
If we assume that the absorbers are Poisson-distributed,
then it is straightforward to show that the effective optical depth
is given by \cite{pmb80}
\bear
&&\tau_{\rm eff}(z,z';\nu) = 
\nline
&&\int_z^{z'} \de z'' \int_0^{\infty} \de N_{\rm HI} 
\f{\del^2 N}{\del z'' \del N_{\rm HI}} 
[1 - {\rm e}^{-N_{\rm HI} \sigma_{\rm abs}(\nu'')}]
\ear
where $[\del^2 N/\del z \del N_{\rm HI}] ~ \de z~\de N_{\rm HI}$ 
is the number of absorbers 
within $(z, z+\de z) $ having column densities
in the range $(N_{\rm HI}, N_{\rm HI} + \de N_{\rm HI})$.

In case of Ly$\alpha$ resonant absorption, we can use the cross section
in (\ref{eq:sigma_alpha}) to calculate $\tau_{\rm eff}$. Since $V$ is a function
which is sharply peaked around $\nu/\nu_{\alpha} = 1$, we can approxiamte
the above integral as
\be
\tau_{\rm eff}(z,z';\nu) = \f{1+z_{\alpha}}{\lambda_{\alpha}}
\int_0^{\infty} \de N_{\rm HI} 
\f{\del^2 N}{\del z_{\alpha} \del N_{\rm HI}} W_{\alpha}(N_{\rm HI}) 
\label{eq:taueff}
\e
where
\be
W_{\alpha}(N_{\rm HI}) \equiv \int \de \lambda'' 
\left[1 - {\rm e}^{-N_{\rm HI} \sigma_{\alpha} V(\lambda_{\alpha}/\lambda''-1)}\right]
\e
is called the ``equivalent width'' of the absorber.

\subsubsection{Continuum absorption}
\label{chap2:contabs}

In case of continuum absorption of radiation by photoionization, the 
cross section is given by
\be
\sigma_{\rm abs}(\nu) = \sigma_{\rm HI}(\nu) \Theta(\nu - \nu_{\rm HI})
\e
where $\sigma_{\rm HI}(\nu)$ is the photoionization cross section and
$\Theta$ is the Heaviside step function taking into account that
only photons with frequencies $\nu > \nu_{\rm HI} = 13.6~{\rm eV}/h_p$
would be absorbed by the photoionization process. The exact
form of $\sigma_{\rm HI}(\nu)$ is rather complicated, however one
can approximate it by a power-law of the form
$\sigma_{\rm HI}(\nu) = \sigma_0 (\nu/\nu_{\rm HI})^{-3}$
where $\sigma_0 = 6.3 \times 10^{-18} {\rm cm}^2$.

Since $\sigma_0 \sim \sigma_{\alpha}$, one can show that the
absorption due to a diffuse IGM in this case too is negligibly small. The only 
significant absorption can be seen in very high density regions which 
have a large fraction of their hydrogen in neutral form. In that case,
we can use the relations obtained for a set of Poisson-distributed
absorbers in the vase of resonant transition. We essentially have an
optical depth of the form (\ref{eq:taueff}), and 
the corresponding mean free path of ionizing photons 
due to these discrete absorbers is found to be
\cite{mhr99,sb03,miralda03}
\bear
\lambda_{\nu}(z) 
&=& \f{c}{H(z) (1+z)}
\nline
&\times&
\left[
\int_0^{\infty} \de N_{\rm HI} 
\f{\del^2 N}{\del z \del N_{\rm HI}} 
[1 - {\rm e}^{-N_{\rm HI} \sigma_{\rm HI}(\nu)}]
\right]^{-1}
\label{eq:lambda_nu}
\ear

At this point, let us introduce the concept of Lyman-limit systems which 
have column densities $N_{\rm HI} > \sigma^{-1}_0
= 1.6 \times 10^{17} {\rm cm}^{-2}$; 
these absorbers contribute a optical depth of unity to the ionizing 
photons. The average distance between these systems is given by
\be
\lambda_{\rm LLS} =  \f{c}{H(z) (1+z)} \left[
\f{\de N_{\rm LLS}}{\de z} 
\right]^{-1}
\label{eq:lambdaLLS}
\e
where $\de N_{\rm LLS}/\de z$ is the redshift distribution of the Lyman-limit
systems
\be
\f{\de N_{\rm LLS}}{\de z}  =  
\int_{1/\sigma_{\rm HI}(\nu_{\rm HI})}^{\infty} \de N_{\rm HI} 
\f{\del^2 N}{\del z \del N_{\rm HI}} 
\e
For the observed distribution $\del^2 N/\del z \del N_{\rm HI} \propto 
N_{\rm HI}^{-1.5}$ \cite{pbcp93}, 
one can show from equations (\ref{eq:lambda_nu})  and 
(\ref{eq:lambdaLLS}) that the mean free path is related
to the distance and redshift distribution Lyman-limit systems as
\be
\lambda_{\nu_{\rm HI}} = \f{\lambda_{\rm LLS}}{\sqrt{\pi}} 
= \f{c}{\sqrt{\pi} H(z) (1+z) ~ \de N_{\rm LLS}/\de z}.
\label{eq:dNLLdz}
\e
The redshift distribution of Lyman-limit systems 
$\de N_{\rm LLS}/\de z$ is a quantity
which has been measured for $2< z < 4.5$ 
by observations of quasar absopriton spectra. Though
the observational constraints are poor, one can still obtain a 
value $\de N_{\rm LLS}/\de z \approx 0.3 (1+z)^{-1.55}$ \cite{smi96}, 
which in turn
gives the mean free path as $\lambda_{\nu_{\rm HI}}/[c/H(z)]
\approx 0.1 [(1+z)/4]^{-2.55}$
Hence the mean free path of ionizing photons
is much smaller
than the horizon size for $z > 3$, which implies that we can use the 
local absorption approximation at these redshifts.

We can summarise the main results of this section as: the post-reionization 
epoch is characterized by a highly ionized IGM as observed by the 
quasar absorption spectra. The IGM is largely transparent to ionizing photons at
these redshifts. However, there exist regions with high column
densities ($N_{\rm HI} > 10^{17} {\rm cm}^{-2} $) 
which are optically thick to the ionizing
radiation; these regions determine the photon mean free path. 
We shall see later how
to use this information to obatin an improved model of the IGM.

\subsection{Pre-overlap epoch}

We now turn our attention towards the IGM in the pre-overlap era.
In this era, the overlap of individual ionized regions is not
complete and hence the IGM is partially ionized. 
So the radiation transfer equation has to be modified 
to account for the multi-phase nature of the IGM. 

Let us define
the volume filling factor of ionized regions to be
$Q_{\rm HII}$; this the fraction of volume that is ionized and 
reionization is said to be complete when $Q_{\rm HII}=1$.
Next, note that the number density of photons present in the background
flux is
\be
n_J(t) = \f{4 \pi}{c} \int_{\nu_{\rm HI}}^{\infty} \de \nu \f{J_{\nu}}{h_p \nu}
\e
Since there is no ionizing flux within the neutral regions (otherwise they
would not remain neutral), 
the photoionization rate per hydrogen atom {\it within the ionized
(HII) regions} is
\be
\Gamma^{\rm II}_{\rm HI} 
= \f{1}{Q_{\rm HII}} 4 \pi \int_{\nu_{\rm HI}}^{\infty} \de \nu 
\f{J_{\nu}}{h_p \nu} \sigma_{\rm HI}(\nu)
\e
where the factor $Q_{\rm HII}^{-1}$ accounts for the fact that the radiation
is limited to a fraction of the total volume. The emmission rate of ionizing
photons per unit volume from sources of emissivity $\epsilon_{\nu}$ is
\be
\dot{n}_{\rm ph} = 
\int_{\nu_{\rm HI}}^{\infty} \de \nu \f{\epsilon_{\nu}}{h_p \nu} 
\e

Then the equation of radiation transfer then becomes
\cite{mhr00,co00}
\bear
\dot{n}_J
&=&  - 3 H(t) n_J
- H(t) \f{4 \pi}{c} \f{J_{\nu_{\rm HI}}}{h_p}
+ \dot{n}_{\rm ph}
\nline
&-& n_{\rm HI}^{\rm II} Q_{\rm HII} \Gamma^{\rm II}_{\rm HI}
- n_{\rm HII}^{\rm II} \f{\de Q_{\rm HII}}{\de t}
\label{eq:radtrans_pre}
\ear
where $n_{\rm HI}^{\rm II}$ and $n_{\rm HII}^{\rm II}$ are the
number densities of neutral and ionized hydrogen within the 
HII regions, respectively.
The first term in the right hand side of equation (\ref{eq:radtrans_pre})
corresponds to the dilution in density due to cosmic expansion while the
second term accounts for the loss of ionizing radiation because of a
photon being redshifted below the ionization edge of hydrogen $\nu_{\rm HI}$.
The third term is essentially the source of ionizing photons. 
The fourth term accounts for the 
loss of photons in ionizing the residual neutral hydrogen within the
ionized regions. The fifth term, which is only relevant for the
pre-overlap stages, accounts for the photons which ionize
hydrogen for the first time and hence increase the filling
factor $Q_{\rm HII}$. For $Q_{\rm HII} = 1$, equation (\ref{eq:radtrans_pre})
reduces to that for the post-reionization phase.

If we now assume that the photons are absorbed locally, then
$\dot{J}, H J  \ll c \epsilon$ and we $J_{\nu}$ is essentially
given by equation (\ref{eq:jnu_mfp}). We can then ignore
terms containing $J$ and $n_J$ in equation (\ref{eq:radtrans_pre}).
This gives a equation describing the evolution
of the filling factor $Q_{\rm HII}$
\be
\f{\de Q_{\rm HII}}{\de t} = 
\f{\dot{n}_{\rm ph}}{n_{\rm HII}^{\rm II}}
- Q_{\rm HII} \Gamma^{\rm II}_{\rm HI}
\e
If we further assume photoionization equilibrium within the ionized
region $\de (n_{\rm HI}^{II} a^3)/\de t \to 0$, then we have from equation 
(\ref{eq:dnhidt_global}) $n_{\rm HI}^{\rm II} \Gamma^{\rm II}_{\rm HI}
= {\cal C} \alpha(T) n_{\rm HII}^{\rm II} n_e^{\rm II}$
and the evolution of $Q_{\rm HII}$ can be written in the form
\cite{mhr99}
\be
\f{\de Q_{\rm HII}}{\de t} = 
\f{\dot{n}_{\rm ph}}{n_{\rm HII}^{\rm II}}
- Q_{\rm HII} ~ {\cal C} \alpha(T) n_e^{\rm II}
\label{eq:dQdt}
\e
In this description, reionization is complete 
when $Q_{\rm HII}=1$ and equation (\ref{eq:dQdt}) cannot be evolved
further on.
Clearly the assumptions of local absorption and photoionization
equilibrium (both of which are reasonably accurate) has given us 
a equation which can be solved once we have a model
for estimating $\epsilon_{\nu}$ and ${\cal C}$. 
Of course, there
is a dependence of the recombination rate coefficient $\alpha(T)$ on
temperature, however that dependence is often ignored while
studying the volume filling factor. In case one is interested
in temperature evolution, one has to solve equation (\ref{eq:dEkindt_global})
taking into account all the heating and cooling processes in the IGM, 
in particular, the photoheating by ionizing photons whose rate is given by
\be
\Gamma_{\rm ph, HI} 
= 4 \pi \int_{\nu_{\rm HI}}^{\infty} \de \nu 
\f{J_{\nu}}{h_p \nu} h_p (\nu - \nu_{\rm HI}) \sigma_{\rm HI}(\nu).
\e

\subsection{Reionization of the inhomogenous IGM}

The description of reionization in the previous section is not adequate
as it does not take into account the inhomogeneities in the IGM
appropriately (except for a clumping factor ${\cal C}$ in the effective
recombination rate). To see this,
consider the post-reionization phase where we know from 
observations that there
exists regions of high density which are neutral; these regions
are being gradually ionized and hence one would ideally like
to write a equation similar to (\ref{eq:dQdt}) for studying the 
post-reionization phase. Since the ionization state
depends on the density, one should have to account for the
density distribution of the IGM.

In order to proceed, first note that the volume filling
factor may not be the appropriate quantity to study for 
evolution of reionization because most of the photons are
consumed in regions with high densities (which might be occupying
a small fraction of volume). 
In other words, if we neglect recomibation for the moment, we have
from equation (\ref{eq:dQdt}) that 
the volume filling factor 
$Q_{\rm HII} = \int \de t~\dot{n}_{\rm ph}/n_H = n_{\rm ph}/n_H$; however, 
in reality the photon to hydrogen ratio should be equal to the ionized
mass fraction $F^M_{\rm HII}$, i.e., $n_{\rm ph}/n_H = F^M_{\rm HII}$.
Hence, we must replace 
the volume filling factor by the mass filling factor in the description
of the previous section, in particular equation (\ref{eq:dQdt}) should 
have the form
\be
\f{\de F^M_{\rm HII}}{\de t} = 
\f{\dot{n}_{\rm ph}}{n_{\rm HII}^{\rm II}}
- F^M_{\rm HII} ~ {\cal C} \alpha(T) n_e^{\rm II}
\e

One can relate $F^M_{\rm HII}$ to the IGM density distribution by using
the fact that regions of lower densities will be ionized first,
and high-density regions will remain neutral for a longer time. The
main reason for this is that the recombination rate (which is $\propto n_H^2$) 
is higher in 
high-density regions where dense gas becomes neutral very quickly. 
If we assume that hydrogen in all regions with overdensities
$\Delta < \Delta_{\rm HII}$ is ionized while the rest is neutral, then
the mass ionized fraction is clearly \cite{mhr00}
\be
F^M_{\rm HII} \equiv F^M(\Delta_{\rm HII}) 
= \int_0^{\Delta_{\rm HII}} \de \Delta ~ \Delta ~ P(\Delta)
\e
where $P(\Delta)$ is the (volume-weighted) density distribution of the IGM.
The term describing the effective recombination rate gets constribution
only from the low desnity regions (high density neutral regions do not 
contribute) and is then given by
\be
\alpha(T) n_e^{\rm II} 
\int_0^{\Delta_{\rm HII}} \de \Delta ~ \Delta^2 ~ P(\Delta)
\equiv \alpha(T) n_e^{\rm II} R(\Delta_{\rm HII}) 
\e
The evolution for the mass ionized fraction is then
\be
\f{\de F^M(\Delta_{\rm HII})}{\de t} = 
\f{\dot{n}_{\rm ph}}{n_{\rm HII}^{\rm II}}
- R(\Delta_{\rm HII}) \alpha(T) n_e^{\rm II}
\label{eq:dfm_dt}
\e
The evolution equation essentially tracks the evolution
of $\Delta_{\rm HII}$ which rises as $F^M(\Delta_{\rm HII})$ increases with
time (i.e., more and more high density regions are getting ionized). Since
the mean free path is determined by the high density regions, one should
be able to relate it to the value of $\Delta_{\rm HII}$ \cite{mhr00}.
It is clear that a photon will be 
able to travel through the low density ionized volume
\be
F_V(\Delta_{\rm HII}) = \int_0^{\Delta_{\rm HII}} \de \Delta ~ P(\Delta)
\e
before being absorbed. When a very high fraction of volume is ionized,  
one can assume that the fraction of volume filled up by the 
high density regions is $1 - F_V$, hence their size is proportional  
to $(1 - F_V)^{1/3}$, and the separation between them along a random 
line of sight will be proportional to $(1 - F_V)^{-2/3}$, 
which, in turn, will determine the 
mean free path. Then one has
\be
\lambda_{\nu}(a) = \f{\lambda_0}{[1 - F_V(\Delta_{\rm HII})]^{2/3}}
\label{eq:lambda_0}
\e
where we can fix $\lambda_0$ by comparing with low redshift observations
like the distribution of Lyman-limit systems [equation (\ref{eq:dNLLdz})].

The situation is slightly more complicated when the ionized regions
are in the pre-overlap stage. At this stage, a
volume fraction $1 - Q_{\rm HII}$ of the universe is completely neutral
(irrespective of the density), while the remaining $Q_{\rm HII}$ fraction of
the volume is occupied by ionized regions. However, within this
ionized volume, the high density regions (with $\Delta > \Delta_{\rm HII}$)
will still be neutral. Once $Q_{\rm HII}$ becomes unity, all regions with
$\Delta < \Delta_{\rm HII}$ are ionized and the rest are neutral; this
can be thought of as the end of reionization.  
The generalization of equation (\ref{eq:dfm_dt}), appropriate for this
description is given by \cite{mhr00,wl03}
\be
\f{\de [Q_{\rm HII} F_M(\Delta_{\rm HII})]}{\de t} = 
\f{\dot{n}_{\rm ph}(z)}{n_{\rm HII}^{\rm II}} 
- Q_{\rm HII} \alpha_R(T) n_e R(\Delta_{\rm HII})
\label{eq:qifm}
\e

Note that there are two unknowns $Q_{\rm HII}$ and 
$F_M(\Delta_{\rm HII})$ in equation (\ref{eq:qifm}) which is 
impossible to solve
it without more assumptions. One assumption which is
usually made is that $\Delta_{\rm HII}$ does not evolve significantly 
with time in the pre-overlap stage, i.e., it is equal to a critical
value $\Delta_c$. This critical density is 
determined from the the mean separation of the 
ionizing sources.
To have some idea about the value of $\Delta_c$, 
two arguments have been put forward in the literature: In the first, 
it is argued that $\Delta_c$ is determined by the distribution of
sources \cite{mhr00,wl03}. When the sources are very numerous, every low-density
region (void)
can be ionized by sources located at the
edges, and hence the overlap of ionized regions can
occur (i.e., $Q_{\rm HII}$
approaches unity) when $\Delta_c \sim 1$ is the characteristic overdensity
of the thin walls separating the voids. For rare and luminous sources,
the mean separation is much larger and hence 
the value of $\Delta_c$ has to be higher before $Q_{\rm HII}$ can be close to
unity. In the second approach, it is assumed that the mean free path is 
determined by
the distance between collapsed objects (which manifest themselves as
Lyman-limit systems) and hence $\Delta_c$ should be similar to the
typical overdensities near the boundaries of the collapsed haloes
\cite{cfo03,cf05}. It usually
turns out to be $\sim 50-60$ depending on the density profile of the halo.
Interestingly, results do not
vary considerably as $\Delta_c$ is varied from $\sim 10$ to $\sim
100$.  Once $\Delta_c$ is fixed, one can follow
the evolution of $Q_{\rm HII}$ until it becomes unity. Following that, we
enter the post-overlap stage, where the situation is well-described by
equation (\ref{eq:dfm_dt}).

Of course, the above descrition is also not fully adequate as 
there will be a dependence on how far the high density 
region is from an ionizing source. A dense region which is very close
to an ionizing source will be ionized quite early compared to, say,
a low-density region which is far away from luminous sources.
However, it has been found that the above description gives a reasonable
analytical description of the reionization process, particularly
for the post-reionization phase. The main advantages in this approach are
(i) it takes into account both the pre-overlap and post-overlap phases
under a single formalism, (ii) once we have some form for the IGM 
density distribution $P(\Delta)$, we can calculate the clumping factor 
and the effective recombination rate self-consistently 
without introducing any extra parameter;
in addition we can also compute the mean free path using one single parameter
($\lambda_0$, which can be fixed by comparing with low-redshift observations).


\section{Modelling of reionization}
\label{chap3}

Given the formalism we have outlined in the previous section, we
can now go forward and discuss some other details involved 
in modelling reionization. 

\subsection{Reionization sources}

The main uncertainty in any reionization model is to identify
the sources. The most natural sources which have been observed to
produce ionizing photons are the star-forming galaxies and quasars. 
Among these, the quasar population is seen to decrease rapidly at $z > 3$
and there is still no evidence of a significant population at higher
redshifts. Hence, the most common sources studied in this area
are the galaxies.

The subject area of formation of galaxies is quite involved in itself
dealing with formation of non-linear structures (haloes and filaments), 
gas cooling and generation of radiation from stars.

\subsubsection{Mass function of collapsed haloes}

The crucial ingredient  for galaxy formation is the 
collapse and virialization of dark matter haloes. 
This can be adequately described by the Press-Schechter 
formalism for most purposes. 
It can be shown that the 
number density of collapsed objects per unit comoving 
volume (which is physical volume divided by $a^3$) 
within a mass range $(M, M+\de  M)$ at an epoch $t$ is given by
\cite{ps74}
\be
\f{\del n(M,t)}{\del M} \de  M = -\sqrt{\f{2}{\pi}} {\rm e}^{-\nu^2/2} 
\f{\bar{\rho}_m}{M} \f{\de \ln \sigma(M)}{\de \ln M}
\f{\nu}{M} \de M,
\label{eq:psnm}
\e
where $\bar{\rho}_m$ is the comoving density of dark matter,
$\sigma(M)$ is defined as 
the rms mass fluctuation at a mass scale~$M$ at $z = 0$, 
$\nu \equiv \delta_c/[D(t) \sigma(M)]$, 
$D(t)$ is the growth factor for linear dark matter perturbations and 
$\delta_c$ is the 
critical overdensity for collapse, usually taken to be equal to 1.69 
for a matter-dominated flat universe $(\Omega_m=1)$. 
This formalism can be extended to calculate 
the comoving number 
density of collapsed objects having mass in the range $(M, M + \de  M)$,
which are formed within the time interval 
$(t_{\rm form}, t_{\rm form} + \de  t_{\rm form})$ 
and observed at a later time $t$ \cite{sasaki94,co00} is given by
\bear
\f{\del^2 n(M,t;t_{\rm form})}{\del M \del t_{\rm form}} 
\de  M \de  t_{\rm form} 
&=& \left.\f{\del^2 n(M,t_{\rm form})}{\del M \del t_{\rm form}}\right|_{\rm form} 
\nline
&\times&
p_{\rm surv}[t|t_{\rm form}] 
~\de  M \de  t_{\rm form} 
\nline
\ear
where $\del^2 n(M,t_{\rm form})/\del M \del t_{\rm form}|_{\rm form}$ is the
formation rate of haloes at $t_{\rm form}$ and 
$p_{\rm surv}[t|t_{\rm form}]$ is their survival
probability till time $t$. Assuming that haloes are destroyed
only when they merge to a halo of higher mass,
both these quantities can be calculated
from the merger rates of haloes. The merger rates can be
calculated  using detailed properties of gaussian random field. 
The quantities can also be calculated in a more simplistic manner 
by assuming that the merger probability is scale invariant; in that case
\cite{sasaki94}
\bear
\left.\f{\del^2 n(M,t)}{\del M \del t}\right|_{\rm form} 
&=& \f{\del^2 n(M,t)}{\del M \del t}
+ \f{\del n(M,t)}{\del M} \f{\dot{D}}{D}
\nline
&=& \f{\del n(M,t)}{\del M} ~\nu^2~ \f{\dot{D}}{D}
\ear
and
\be 
p_{\rm surv}[t|t_{\rm form}] = \f{D(t)}{D(t_{\rm form})}.
\e

\subsubsection{Star formation rate}

If these dark matter 
haloes are massive enough to form huge potential wells, 
the baryonic gas will simply fall into those wells. 
As the gas begins to settle into the dark matter haloes, 
mergers will heat it up to the virial temperature 
via shocks. 
However, to form galaxies, the gas has to 
dissipate its thermal energy and cool. 
If the gas contains only atomic hydrogen, it is unable 
to cool at temperatures lower than $10^4$ K because the atomic 
hydrogen recombines and cannot 
be ionized by collisions. The gas can cool effectively 
for much lower temperatures in the presence of 
molecules -- however, 
it not at all clear whether there are sufficient amount 
of molecules  present in the gas at high redshifts 
\cite{har00,rgs01,gb01}. 
Hence the lower mass cutoff for the 
haloes which can host star formation will be
decided by the cooling efficiency of the baryons. 

Let $\dot{M}_*(M,t,t_{\rm form})$ denote the rate of star formation
at time $t$ within a halo of mass $M$ which has formed at $t_{\rm form}$.
Then we can write the cosmic SFR per unit 
volume at a time $t$,
\bear
\dot{\rho}_*(t) &=& \f{1}{a^3(t)} \int_0^t \de  t_{\rm form} 
\int_{M_{\rm min}(t)}^{\infty} \de  M' \dot{M}_*(M',t,t_{\rm form}) 
\nline
&\times& \f{\del^2 n(M',t;t_{\rm form})}{\del M' \del t_{\rm form}} 
\ear
where the $a^{-3}$ is included to covert from comoving to physical volume.
The lower mass cutoff $M_{\rm min}(t)$ at a given epoch is decided by the 
cooling criteria as explained above. However, once reionization starts and
regions are reheated by photoheating, the value of $M_{\rm min}(t)$ is set
by the 
photoionization temperature $\simeq 10^4$ K. This can further suppress
star formation in low mass haloes and is known as radiative feedback. 
We shall discuss this later in this section.

The form of $\dot{M}_*(M,t,t_{\rm form})$ contains information about
various cooling and star-forming processes and hence is quite complex
to deal with. 
It can be obtained from semi-analytical modelling of galaxy formation
\cite{sp99} or constrained from observations of galaxy luminosity function
\cite{sss07}. 
A very simple assumption that is usually made for modelling
reionization is that the
duration of star formation is much less than the Hubble time 
$H^{-1}(t)$ which is motivated by the fact that most of the ionizing
radiation is produced by hot stars which have shorter lifetime. In that
case, $\dot{M}_*(M,t,t_{\rm form})$ can be approximated
as
\be
\dot{M}_*(M,t,t_{\rm form}) \approx M_* \delta_D(t - t_{\rm form})
= f_* \f{\bar{\rho}_b}{\bar{\rho}_m} M \delta_D(t - t_{\rm form})
\e
where $\bar{\rho}_b/\bar{\rho}_m M$ is the mass of baryons within the halo and
$f_*$ is the fraction of baryonic mass which has been converted
into stars. The cosmic SFR per comoving volume is then
\be
\dot{\rho}_*(t) =  \f{1}{a^3(t)} \f{\bar{\rho}_b}{\bar{\rho}_m}
\int_{M_{\rm min}(t)}^{\infty} \de  M' ~ f_* ~ M'
\left.\f{\del^2 n(M',t)}{\del M' \del t}\right|_{\rm form} 
\e
One should keep in mind that many details of the star-formation process
has been encoded within a single parameter $f_*$. This should, in principle,
be a function of both halo mass $M$ and time $t$. However, it is not clear 
at all what the exact dependencies should be. Given such uncertainties, it
is usual to take it as a constant.

In addition, one finds that the merger rate of haloes at high redshifts
is much less than the formation rate (which follows 
if $\nu \gg 1$), hence
$\del^2 n(M,t)/\del M \del t|_{\rm form} \approx \del^2 n(M,t)/\del M \del t$.
Then, one can write the SFR in terms of the fraction of collapsed mass
in haloes more massive than $M_{\rm min}(t)$
\bear
f_{\rm coll}(t) &=& \f{1}{\bar{\rho}_m} 
\int_{M_{\rm min}(t)}^{\infty} \de  M' ~ M' \f{\del n(M',t)}{\del M'}
\nline
&=& {\rm erfc}\left[\f{\delta_c}{\sqrt{2}~D(t) \sigma(M_{\rm min})}\right]
\label{eq:fcoll}
\ear
as
\be
\dot{\rho}_*(t) = f_* \f{\bar{\rho}_b}{a^3(t)} \f{\de f_{\rm coll}(t)}{\de t}
\e

\subsubsection{Production of ionizing photons}

Given the SFR, we can calculate the emissivity of galaxies, or 
equivalently 
the rate of ionizing photons in the IGM 
per unit volume per unit frequency range:
\be
\dot{n}_{\nu, {\rm ph}}(t) = f_{\rm esc} \left[\f{\de N_{\nu}}{\de M_*}\right]
\dot{\rho}_*(t)
\e
where $\de N_{\nu}/\de M_*$ gives the number of photons emitted per  
frequency range per unit mass of stars and $f_{\rm esc}$ 
is the fraction of ionizing photons which escape from 
the star forming haloes into the IGM. The emissivity is simply
$\epsilon_{\nu} = h_p \nu \dot{n}_{\nu, {\rm ph}}$.

Given the spectra of stars of different 
masses in a galaxy, and their Initial Mass Function (IMF), 
$\de N_{\nu}/\de M_*$ 
can be computed in a straightforward way using ``population synthesis'' codes 
\cite{lsg++99,bc03}.
The IMF and spectra will depend on the details of star formation 
(burst formation or continuous) and metallicity. In fact, it is possible that
there are more than one population of stellar sources which have different
$\de N_{\nu}/\de M_*$. For example, 
there are strong indications, both from numerical simulations and analytical
arguments \cite{bkl01}, 
that the first generation stars were metal-free, and hence
massive, with a very different kind of IMF and spectra than the stars 
we observe today \cite{schaerer02}; they are known as the Population III 
(PopIII) stars.

A fraction of photons produced in a galaxy would be consumed in ionizing the
neutral matter within the galaxy itself. Hence only a fraction of 
photons escape into the IGM and is available for reionization, which is
encoded in the parameter $f_{\rm esc}$. This parameter is again not very
well modelled \cite{rs00,wl00,gkc08,wc09} 
and its observed value is also quite uncertain \cite{spa01,sspae06,iid06,cpg07}; typical values
assumed are $\sim 0.1$. It is most likely
$f_{\rm esc}$, like $f_*$, is also 
a function of halo mass and the time of halo formation, 
however since the dependences are not well understood, it is taken to 
be a constant.

The total number of ionizing photons is then obtained by integrating the
above quantity over all energies above the ionization threshold:
\be
\dot{n}_{\rm ph}(t) = \int_{\nu_{\rm HI}}^{\infty} \de \nu ~ \dot{n}_{\nu, {\rm ph}}(t)
= N_{\rm ion} n_b \f{\de f_{\rm coll}(t)}{\de t}
\label{eq:ndot_fcoll}
\e
where $n_b$ is the total baryonic number density in the IGM (equal to $n_H$ if 
we ignore the presence of helium) and 
\be
N_{\rm ion} \equiv f_* f_{\rm esc} m_p \int_{\nu_{\rm HI}}^{\infty} \de \nu
\left[\f{\de N_{\nu}}{\de M_*}\right]
\e
is the number of photons entering the IGM
per baryon in collapsed objects  \cite{wl07}.
In case there are more than one population of stars, 
one has to use different values of $N_{\rm ion}$ for the
different populations.

\subsubsection{Feedback processes}
\label{chap3:feedback}

The moment there is formation of stars and other luminous bodies,
they start to affect the subsequent formation of structures -- 
this is known as feedback \cite{cf05a}. The process is intrinsically non-linear and
hence quite complex to model.
The feedback processes can be categorized roughly into three
categories.

The first of them is the radiative feedback which is associated
with the radiation from first stars which can heat up the
medium and can photoionize atoms and/or
photodissociate molecules. 
Once the first galaxies form stars, their radiation
will ionize and heat the surrounding medium, increasing the mass scale 
(often referred to as the {\it filtering mass} \cite{gh98}) above which baryons 
can collapse 
into haloes within those regions. The minimum mass of haloes which 
are able to cool is thus much higher in ionized regions than in neutral ones. 
Since the IGM is multi-phase in the pre-overlap phase, 
one needs to take into account the heating of the ionized regions right from 
the beginning of reionization. In principle, this can be done self-consistently
from the evolution of the temperature of the ionized region 
in equation (\ref{eq:dEkindt_global}).

The low mass haloes can be subjected to mechanical feedback too,
which is mostly due to energy injection via supernova explosion
and winds. This can expel the gas from the halo and suppress
star formation. As in the case for radiative feedback, 
one can parametrize this through the
minimum mass parameter $M_{\rm min}(t)$.

Finally, we also have chemical feedback where the stars expel
metals into the medium and hence change the chemical composition.
This would mean that the subsequent formation of stars could be
in a completely different environment and hence the nature
of stars would be highly different.

\subsubsection{Quasars}

Besides the stellar sources, a population of
accreting black holes or quasars are also known to produce significant
amount of ionizing radiation. Hence it is possible that they too have
contributed to reionization. The fraction of their contribution would
depend on the number of quasars produced at a particular redshift.
Observationally, the luminosity function of quasars is quite well-probed
till a redshift $z \sim 6$ \cite{rsf++06}. 
It turns out that the population
peaks around $z \approx 3$ and decreases for higher redshifts. Hence
their contribution at higher redshifts is highly debated.

The difference between reionization by stellar sources and quasars lie
in the fact that quasars produce significant number of high energy photons
compared to stars. This would imply that quasars can contribute
significantly to double-reionization of helium (which requires
photons with energies $> 54.4$ eV, not seen in galaxies). In addition,
quasars produce significant amount of X-ray radiation. 
Since the absorption cross section of neutral hydrogen varies 
with frequency approximately
as $\nu^{-3}$, the mean free path for photons with high energies 
would be very large. A simple calculation will show that for photons
with energies above 100--200 eV, the mean free path would be larger
than the typical separation between collapsed structures \cite{mrvho04}
(the details
would depend upon the redshift and exact description of collapsed 
haloes). These photons would not be associated with any particular
source at the moment when they are absorbed, and thus would
ionize the IGM in a more homogeneous manner (as opposed to the 
overlapping bubble picture for UV sources).

\subsection{Illustration of a semi-analytical model}
\label{chap3_cfmodel}

\begin{figure*}
\rotatebox{270}{\resizebox{0.5\textwidth}{!}{\includegraphics{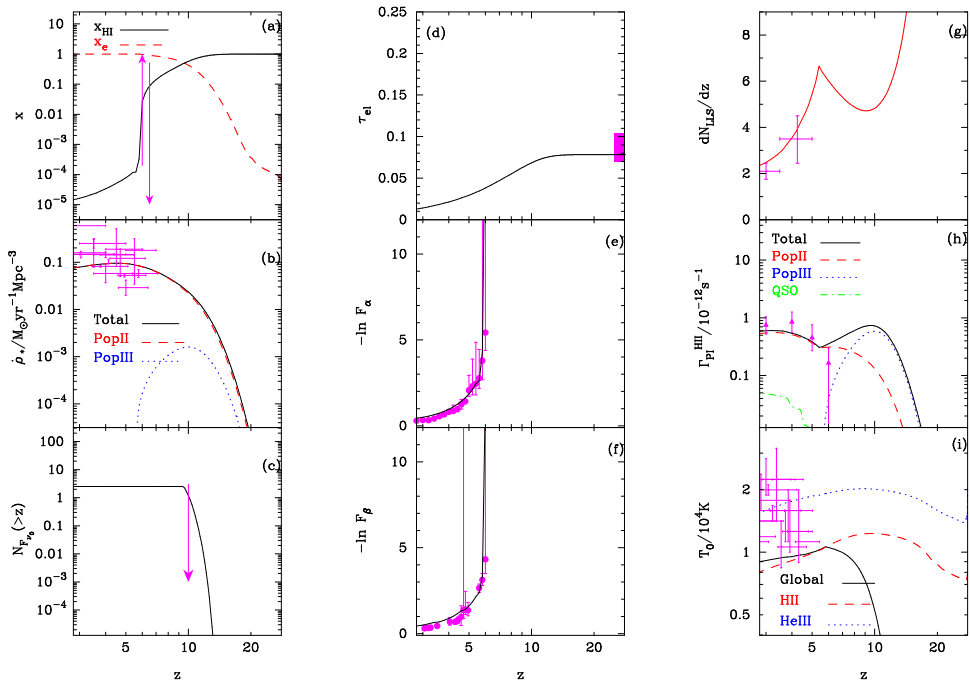}}}
\caption{Comparison of analytical model predictions with observations for
the best-fit model. The different panels indicate:
(a) The volume-averaged neutral hydrogen fraction $x_{\rm HI}$, with 
observational
lower limit from quasar absorption lines at $z =6$ and upper limit from
Ly$\alpha$ emitters at $z = 6.5$ (shown with arrows). In addition, the 
ionized fraction $x_e$ is shown by the dashed line.
(b) SFR $\dot{\rho}_*$ for different stellar populations. 
(c) The number of source
counts above a given redshift, 
with the observational upper limit from NICMOS HUDF shown by the
arrow. The contribution to the source count is zero at low 
redshifts because of the 
J-dropout selection criterion.
(d) Electron scattering optical depth, with  observational constraint from
WMAP 3-year data release. (e) Ly$\alpha$ effective optical
depth.
(f) Ly$\beta$ effective optical
depth.
(g) Evolution of Lyman-limit systems. (h) Photoionization
rates for hydrogen, with estimates from numerical simulations (shown
by points with error-bars; \cite{bhvs05}). (i) Temperature of the mean
density IGM.
}
\label{fig:bf}
\end{figure*}

The physics described above in the preceding sections can all
be combined
to construct a semi-analytical model for studying the thermal
and ionization history of the IGM. We shall give an explicit
example of one such model \cite{cf05,cf06b} whose main features are:  
The model accounts for IGM inhomogeneities by adopting a lognormal 
distribution for $P(\Delta)$; 
reionization is said to be complete once all the low-density regions (say, with overdensities $\Delta < \Delta_c \sim 60$) are ionized. 
The ionization and thermal histories 
of neutral, HII and HeIII regions are followed 
simultaneously and self-consistently, treating the IGM as a 
multi-phase medium. 
Three types of reionization sources have been assumed: (i) metal-free (i.e. PopIII) stars having a Salpeter IMF in the mass 
range $1 - 100 M_{\odot}$: they dominate the photoionization rate at high redshifts; (ii) PopII stars with sub-solar metallicities 
also having a Salpeter IMF in the mass range $1 - 100 M_{\odot}$; 
(iii) quasars, which are 
significant sources of hard photons at $z \lesssim 6$; they have negligible effects on the IGM at higher redshifts.

As discussed earlier, reionization is accompanied by various feedback processes, which can affect subsequent star formation. In this model, radiative 
feedback is computed self-consistently from the evolution of the 
thermal properties of the IGM. Furthermore, 
the chemical feedback inducing the PopIII $\rightarrow$ PopII transition is implemented using a merger-tree ``genetic'' 
approach which determines the termination of PopIII 
star formation in a metal-enriched halo \cite{ssfc06}.

The predictions of the model are compared with a wide range of observational data sets, namely, (i) redshift evolution of Lyman-limit absorption 
systems $\de N_{\rm LLS}/\de z$ \cite{smih94}, (ii) the effective
optical depths $\tau_{\rm eff} \equiv -\ln F$ for Ly$\alpha$ and 
Ly$\beta$ absorption in the IGM \cite{songaila04}, (iii) 
electron scattering optical depth 
$\tau_{\rm el}= \sigma_T c \int \de t ~ n_e$ (where $\sigma_T$ is the
Thomson scattering cross section) 
as measured from CMBR experiments \cite{sbd++07}, 
(iv) temperature of the mean intergalactic gas \cite{stle99}, (v)  
cosmic star formation history $\dot{\rho}_*$
\cite{nchos04}, and (vi) source number counts at $z 
\approx 10$ from NICMOS HUDF \cite{bitf05}.

The data constrain the reionization scenario quite tightly. We find that hydrogen reionization starts at $z \approx 15$ driven by metal-free 
(PopIII) stars, and it is 90 per cent complete by $z \approx 8$.  
After a rapid initial phase, the growth of the
volume filled by ionized regions slows down at $z \lesssim 10$ due to the 
combined action of chemical and radiative feedback,  
making reionization a considerably extended process completing only at 
$z \approx 6$. The number of photons per hydrogen at the end
of reionization at $z \approx 6$ is only a few, which implies that
reionization occurred in a ``photon-starved'' manner \cite{bh07}.

\subsection{Clustering of sources}

The formalism described till now works quite well for studying
global properties of reionization. However, it is not
adequate for studying the details of overlap
of ionized regions in the pre-reionization epoch. The
shapes and distribution of ionized regions are crucially
dependent on the distribution of sources, in particular, their
clustering and also on the density structure of the IGM. For example,
if the sources are highly clustered, then it is expected that the
overlap of the ionized bubble for nearby sources would be much earlier
than what is expected from a random distribution of sources.

Modelling of reionization to such details have become important 
because of the 21cm observations which are likely
to be available in near future. Before discussing the
theoretical model, let us briefly outline the
motivation behind 21cm experiments.

\subsubsection{21cm observations}

Perhaps the most promising 
prospect of detecting the fluctuations in the neutral 
hydrogen density during the (pre-)reionization era is through 
the 21 cm emission
experiments \cite{fob06}, some of which are
already taking data (GMRT \footnote{http://www.gmrt.ncra.tifr.res.in}, 
21CMA \footnote{http://web.phys.cmu.edu/$\sim$past/}), and some
are expected in future (MWA \footnote{http://www.haystack.mit.edu/arrays/MWA}, 
LOFAR \footnote{http://www.lofar.org}, 
SKA \footnote{ http://www.skatelescope.org/}). 
The basic principle which is central to these experiments is the 
neutral hydrogen hyperfine transition line at a rest wavelength of 
21 cm. This line, when redshifted, is observable
in radio frequencies ($\sim 150$ MHz for $z \sim 10$)
as a brightness temperature:
\begin{equation}
\delta T_b(z, {\bf \hat{n}}) = \frac{T_S - T_{\rm CMB}}{1+z} 
~ \frac{3 c^3 \bar A_{10} n_{\rm HI}(z,{\bf \hat{n}}) (1+z)^3}
{16 k_{\rm boltz} \nu_0^2 T_S H(z)}
\end{equation}
where $T_S$ is the spin temperature of the gas, $T_{\rm CMB} = 2.76 (1 + z)~ {\rm K}$ is the CMBR temperature, $A_{10}$ is the Einstein coefficient
and $\nu_0 = 1420$ MHz is the rest frequency of the hyperfine line.

The observability of this brightness temperature against the 
CMBR background will depend on the relative values 
of $T_S$ and $T_{\rm CMB}$. Depending on which processes
dominate at different epochs, $T_S$ will couple 
either to radiation ($T_{\rm CMB}$) or to matter (determined
by the kinetic temperature $T_k$)  \cite{pl08}.
Almost in all models of reionization, the most interesting phase
for observing the 21 cm radiation is $6 \lesssim z \lesssim 20$. This 
is the phase where the IGM is suitably heated to temperatures
much higher than CMBR (mostly due to X-ray heating \cite{cm04}) thus making
it observable in emission. 
In that case, we have $\delta T_b \propto n_{\rm HI}/n_H$, which means
that the observations would directly probe the neutral hydrogen
density in the Universe.
Furthermore, this is the era when the 
bubble-overlapping phase is most active, and there is substantial
neutral hydrogen to generate a strong enough signal. At low
redshifts, after the IGM is reionized, $n_{\rm HI}$ falls by orders
of magnitude and the 21 cm signal vanishes except in the high density
neutral regions. Since the observations directly probe the
neutral hydrogen density, one can use it to probe the detailed
topology of the ionized regions in the pre-overlap phase. It is
therefore essential to model the clustering of the sources 
accurately so as to predict the reionization topology.

\subsubsection{Models of source clustering}

There have been various approaches to account for the clustering of sources, 
most of them using the properties of the gaussian random field in terms of
the extended Press-Schechter formalism. We have written
equation (\ref{eq:dQdt}) in terms of globally-averaged quantities; now
let us write it in a slightly different form where the
averaging is done over a spherical region of radius $R$ which has
a density contrast $\delta$ (linearly extrapolated to present
epoch). Then \cite{wl07,wm07,gw08}
\be
\f{\de \la Q_{\rm HII} \ra_{\delta, R}}{\de t} = N_{\rm ion}
\f{\de \la f_{\rm coll} \ra_{\delta, R}}{\de t}
- \la Q_{\rm HII} \ra_{\delta, R} ~ {\cal C} \alpha(T) n_H (1 + \delta)
\label{eq:dQ_Rdt}
\e
where we have assumed that reionization is primarily
driven by galaxies and have used equation (\ref{eq:ndot_fcoll}) to write 
$\dot{n}_{\rm ph}$ in terms of $\dot{f}_{\rm coll}$. 
The above equation can be solved for a given $\{\delta, R\}$ if we know the form
of $\la f_{\rm coll} \ra_{\delta, R}$. It turns out that one has a simple
generalization of equation (\ref{eq:fcoll}) which encodes the clustering
of sources at different scales and is given by \cite{bcek91}
\be
\la f_{\rm coll} \ra_{\delta, R} = 
{\rm erfc}\left[\f{\delta_c/D(t) - \delta}
{\sqrt{2 [\sigma^2(M_{\rm min})-\sigma^2(M)]}}\right]
\e
where $M = 4 \pi R^3 \bar{\rho}_m/3$ is the mass scale corresponding to $R$.
The evolution of $\la Q_{\rm HII} \ra_{\delta, R}$ can be used for
studying the filling fraction of ionized regions within the
IGM on various scales $R$ as a function of overdensity $\delta$. 
In typical scenarios, 
this model predicts that
reionization is driven by overlap of individual ionized regions 
around clustered sources residing in overdense regions of the Universe.
This leads to an ``inside-out'' scenario of reionization where, on average,
high-density regions are ionized first.

In a somewhat similar but slightly different approach, one can obtain
the size distribution of the ionized regions at a given epoch. If we
integrate equation (\ref{eq:dQ_Rdt}) upto a certain time $t$, we can
write the filling factor as
\be
\la Q_{\rm HII} \ra_{\delta, R} = \f{N_{\rm ion} 
\la f_{\rm coll} \ra_{\delta, R}}{1 + N_{\rm rec}}
\e
where $\la Q_{\rm HII} \ra_{\delta, R} ~ N_{\rm rec} = \int^t \de t' ~ \la Q_{\rm HII} \ra_{\delta, R}~ {\cal C}~ \alpha(T) n_H (1 + \delta)$
is the number of recombinations within the region over the history of the IGM.
Hence, the condition for the region to be fully ionized 
$\la Q_{\rm HII} \ra_{\delta, R} \geq 1$ is given by a condition
on the collapsed faction \cite{fzh04b,mf07,mlz+07}
\be
\la f_{\rm coll} \ra_{\delta, R} \geq N_{\rm ion}^{-1} (1 + N_{\rm rec})
\e
The condition for a region to be ``self-ionized'' can be converted
into a condition in terms of the density contrast $\delta$. The problem
then is very similar to the problem of collapse of haloes where it is studied
whether the density averaged over a spherical volume exceeds a 
critical value (``barrier'') \cite{st02}. The only difference here is that
the barrier is much complex than the collapse of halo problem. 
One can approximate the barrier with a linear one and then write a distribution
of sizes similar to equation (\ref{eq:psnm}). The results obtained
from this approach is very similar to the one obtained from 
the previous one that reionization at large scales
proceeds in a ``inside-out'' fashion.

Both the approaches described above have also been extended to
simulation boxes and used for making mock maps of the neutral hydrogen
distribution which is extremely useful for 21cm observations.
These methods essentially give a semi-analytic (or semi-numeric) approach
to deal with the radiative transfer problem and can be used
for making maps using much less computing power.

\subsubsection{Recombination and self-shielding}

\begin{figure}
\rotatebox{270}{\resizebox{0.15\textwidth}{!}{\includegraphics{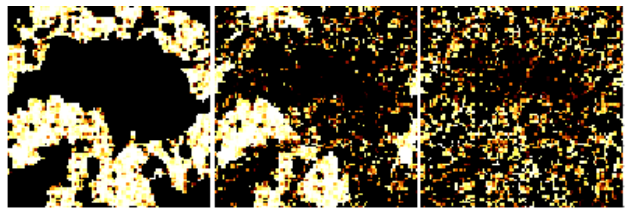}}}
\caption{Ionization maps  for a IGM with ionized hydrogen fraction of 75\%
for models with different assumptions regarding
the spatial distribution  of sinks and sources of ionizing radiation
\cite{chr09}.
The black regions represent ionized ones.
The left panel considers only homogenous recombination while the middle and 
right
panels consider different levels of self-shielding.
 The thickness of the slice shown is
1$h^{-1}$Mpc.}
\label{fig:plot_comparerec}
\end{figure}

In the above approaches, recombinations in the IGM an be accounted
for only by averaging over a spherical region.
In reality, even if a given spherical region contains
enough  photons to self-ionize, the high-density clumps within the
region will remain neutral for  a longer period because of their high
recombination rate and thus alter the nature of the  ionization
field. A  simple prescription to describe the presence of such neutral
clumps  by assuming that regions with overdensities above a critical
value ($\Delta > \Delta_c$)  remain neutral as we discussed earlier. 
Unfortunately this is also  not fully
appropriate  as many of the high-density regions are expected to harbour
ionizing sources.  Whether a region remains neutral will depend on
two competing factors, the local density (which determines the
recombination rate) and  the proximity to ionizing sources (which
determines the number of photons available). It is thus important to
include a realistic spatial distribution of  recombinations into the
formalisms for making ionization maps.

A possible way of modelling the  recombinations in high-density
regions is to use a  self-shielding criterion.  In order to be
ionized, a given point should satisfy the condition that it cannot
remain self-shielded, i.e.,  it should not be a Lyman-limit system. In terms
of the column density of the system, the condition can be written as
\cite{chr09}
\be 
N_{\rm HI} \equiv n_{\rm HI}({\bf x}) L({\bf x}) \sigma_{\rm HI} \leq 1, 
\e 
where $n_{\rm HI}({\bf x})$ is the number density of neutral
hydrogen at the given point and 
$L({\bf x})$ is the size of the of the absorber. It turns out that the 
topology
of reionization is very different if such self-shielding is taken into
account particularly when reionization is extended and photon-starved
as can be seen from Figure \ref{fig:plot_comparerec}.
Initially reionization proceeds inside-out with the
high density regions hosting  the sources of ionizing sources becoming
ionized first. In the later stages, the
high density regions which are far from sources remain neutral and  
reionization proceeds
deep into the underdense regions before slowly  evaporating  denser
regions. Such models can, in principle, be constrained by the first generation
of 21 cm experiments.


\section{Current status and Future} 
\label{chap4}

In this section, let us review the current status of various approaches
to studying reionization and their future prospects.

\subsection{Simulations}

Though the analytical studies mentioned above allow us to develop a
good understanding of the different processes involved in reionization, they
can take into account the physical processes only in some approximate
sense. In fact, a detailed and complete description of reionization would
require locating the ionizing sources, resolving the inhomogeneities 
in the IGM, following the scattering processes through detailed radiative
transfer, and so on. Numerical simulations, in spite of their
limitations, have been of immense 
importance in these areas.

Since the ionizing photons 
during early stages of 
reionization mostly originate from smaller haloes which 
are far more numerous than the larger galaxies at high redshifts.
The need to resolve such small
structures requires the simulation boxes to have high enough resolution. On
the other hand, these ionizing sources were strongly clustered at
high redshifts and, as a consequence, the ionized regions they created
are expected to overlap and grow to very large sizes, 
reaching upto tens of Mpc 
\cite{bl04,fo05,cen05}. As already discussed, the many orders of magnitude
difference between these length scales demand extremely high
computing power from any simulations designed to study early structure 
formation from the point of view of reionization.

To simulate reionization, one usually runs a N-body
simulation (either dark matter only or including baryons) to
generate the large-scale density field, identifies
haloes within the density field and assign ionizing photons
to the haloes using a assumption like equation (\ref{eq:ndot_fcoll}).
It turns out that the most difficult step 
is to solve the radiative transfer equation and study the growth of ionized
regions.
In principle, one could solve equation (\ref{eq:radtrans_local}) directly for the intensity 
at every point in the seven-dimensional $(t,{\bf x}, {\bf n},\nu)$ space, given 
the absorption coefficient and the emissivity. 
However, the high dimensionality of the problem makes the solution 
of the complete radiative transfer equation well beyond our capabilities, 
particularly since we do not have any obvious symmetries in the
problem and often need high spatial and angular resolution in 
cosmological simulations. 
Hence, the approach to the problem has been to use different 
numerical schemes and approximations, like ray-tracing \cite{anm99,rs99,sah01,cen02,rnas02,sir04,bmw04,isr05,impmsa06},  Monte Carlo methods \cite{cfmr01,mfc03,mck09}, 
local depth approximation \cite{go97} and others \cite{ps08}.
At present, most of the simulations do not have enough resolution to 
reliably identify the low mass $\sim 10^8 M_{\odot}$ sources which were
probably responsible for early stages of reionization. Also, there are
difficulties in resolving the small scale structures which 
contribute significantly to the clumpiness in the IGM and 
hence extend the reionization process.

\subsection{Various observational probes}

Finally, we review certain observations which shape our 
understanding of reionization.

\subsubsection{Absorption spectra of high redshift sources}

We have already discussed that the primary evidence for reionization
comes from absorption spectra of quasars (Ly$\alpha$ forest) at $z < 6$. 
We have also discussed that the effective optical depth of Ly$\alpha$
photons becomes
significantly large at $z \gtrsim 6$ implying
regions with high transmission in the Ly$\alpha$ forest 
becoming rare at high redshifts \cite{fsb++06,wdr++09}.
Therefore 
the standard methods of analyzing the Ly$\alpha$ forest (like the 
probability distribution function and power spectrum) are 
not very effective. 
Amongst alternate methods, one can use the the distribution 
of dark gaps \cite{croft98,sc02} which are defined as contiguous regions
of the spectrum having an optical depth above a threshold 
value \cite{sc02,fsb++06}. 
It has been found that the current observations constrain the
neutral hydrogen fraction $x_{\rm HI} < 0.36$ at $z=6.3$ \cite{gffc08}.
It is expected that the SDSS and Palomar-Quest survey \cite{ldc++05}
would detect $\sim 30$ quasars at
these redshifts within the next few years and hence we expect 
robust conclusions from such studies in very near future.

Like quasars, one can also use absorption spectra of other 
high redshift energetic sources like gamma ray bursts (GRBs)
and supernovae.  In fact, analyses using the damping wing 
effects of the Voigt profile
have been already performed on the GRB detected at a
redshift $z = 6.3$, and the wing shape is well-fit
by a neutral fraction $x_{\rm HI}  < 0.17$ \cite{tkk++06}. 
The dark gap width distribution gives a similar constraint 
$x_{\rm HI} = (6.4 \pm 0.3) \times 10^{-5}$ \cite{gsfc08}.
In order to obtain more stringent limits on reionization, 
it is important to increase the sample size of $z >6$ GRBs.

In addition to hydrogen reionization, the 
Ly$\alpha$ forest in the quasar absorption lines
at $z \approx 3$ can also be used for studying reionization of 
singly ionized helium to doubly ionized state (the reionization
of neutral helium to singly ionized state follows hydrogen
for almost all types of sources). The helium reionization coincides
with the rise in quasar population at $z \sim 3$ and it effects
the thermal history of the IGM at these redshifts. However, there
are various aspects of the observation that are not well understood
and requires much detailed modelling of helium reionization
\cite{gnbos05,bvkhc08,fo08,bof08,mlz+09}.

\subsubsection{CMBR observations}

As we have discussed already, the first evidence for an early 
reionization epoch came from the CMBR polarization data. This 
data is going to be much more precise in future with 
experiments like PLANCK, and is expected to improve the 
constraints on $\tau_{\rm el}$. 
With improved statistical errors, 
it might be possible to distinguish between different evolutions
of the ionized fraction, particularly with E-mode polarization 
auto-correlation, as is found
from theoretical calculations \cite{hhkk03,bpsscf08}.
An alternative option to probe reionization through CMBR 
is through the small scale observations of temperature anisotropies.
It has been well known that the scattering of the CMBR photons 
by the bulk motion of the electrons in clusters gives rise to a signal
at large multipoles $\ell \sim 1000$, 
known as the kinetic Sunyaev Zeldovich (SZ) effect.
Such a signal can also originate from the 
fluctuations in the distribution of free electrons 
arising from cosmic reionization. 
It turns out that for reionization, the signal is dominated
by the patchiness in the $n_e$-distribution. 
Now, in most scenarios of reionization, it is expected that 
the distribution of neutral hydrogen would be quite 
patchy in the pre-overlap era, with the ionized hydrogen 
mostly contained within isolated bubbles. 
The amplitude of this signal is significant around 
$\ell \sim 1000$ and is usually comparable to or
greater than the signal arising from standard kinetic SZ effect. 
Theoretical estimates of the signal have been performed 
for various reionization scenarios, and it has been predicted that
the experiment can be used for constraining reionization history
\cite{schkm03,mfhzz05}. 
Also, it is possible to have an idea about the nature of reionization
sources, as the signal from UV sources, X-ray sources and 
decaying particles are quite different. With multi-frequency 
experiments like 
Atacama Cosmology Telescope (ACT)\footnote{http://www.hep.upenn.edu/act/} 
and South Pole Telescope (SPT)\footnote{http://spt.uchicago.edu/}
coming up in near future, this promises to 
put strong constraints on the reionization scenarios.

\subsubsection{Ly$\alpha$ emitters}

In recent years, a number of groups have studied star-forming galaxies at
$z \sim 6 - 7$, and measurements of the Ly$\alpha$ emission line luminosity
function evolution provide another useful observational constraint
\cite{mr04,sye++05}.
While the quasar absorption
spectra probe the neutral hydrogen fraction regime $x_{\rm HI} \leq 0.01$,
this method is sensitive to the range $x_{\rm HI} \sim 0.1 - 1.0$.
Ly$\alpha$ emission from galaxies is expected to be 
suppressed at redshifts beyond reionization because of the 
absorption due to neutral hydrogen, which clearly 
affects the evolution of the luminosity function of such 
Ly$\alpha$ emitters
at high redshifts \cite{hc05,mr04,fhz04}.
Thus a comparison of the luminosity functions
at different redshifts could be used for constraining the reionization.
Through a simple analysis, it was found that the luminosity functions 
at $z = 5.7$  and $z = 6.5$ are statistically consistent with one another
thus implying that reionization was largely complete at $z \approx 6.5$.
More sophisticated calculations on the evolution of the luminosity function of Ly$\alpha$
emitters \cite{mr04,fzh06,hc05} suggest that the neutral fraction
of hydrogen at $z=6.5$ should be less than 50 per cent
\cite{mr06}.
Unfortunately, 
the analysis of the  Ly$\alpha$ emitters at high redshifts is complicated
by various factors like the velocity of the sources with respect to the 
surrounding IGM, the density distribution and the size of ionized regions 
around the sources and the clustering of sources.
It is thus extremely important to have detailed models of 
Ly$\alpha$ emitting galaxies in order to use them for constraining
reionization.

\subsubsection{Sources of reionization}

As we discussed earlier, a major challenge 
in our understanding of reionization depends on our knowledge
of the sources, particularly at high redshifts. 
As we understand at present, neither the bright
$z> 6$ quasars discovered by the SDSS group \cite{fsr++05} nor the
faint ones detected in X-ray observations \cite{bcc++03} 
produce enough photons to reionize the IGM. 
The discovery of star-forming galaxies
at $z>6.5$ \cite{hcm++02,ktk++03,kesr04}
has resulted in speculation that
early galaxies produce bulk of the ionizing photons for reionization. 
Unfortunately, there are significant uncertainties in constraining
the amount of ionizing radiation at these redshifts because the bulk
of ionizing photons could be produced by faint sources which are
beyond the present sensitivities.
In fact, some models have predicted that the $z > 7$  
sources identified in these surveys are relatively massive  
$(M \approx 10^9 M_{\odot}$)  and rare objects 
which are only marginally ($\approx 1\%$) contributing 
to the reionization photon budget \cite{cf07}.
A much better prospect of detecting these sources would be 
through the Ultra-Deep Imaging Survey using the future telescope JWST.

\subsubsection{21cm experiments}

We have already discussed the basic theory behind 
detecting the fluctuations in the neutral 
hydrogen density through 
the 21 cm emission
experiments. There are essentially two complementary
approaches to studying reionization using 21 cm signal. The first
one is through global statistical properties of the 
neutral hydrogen signal, like the power spectrum
\cite{mh04b,zfh04,abp05,ba04,sethi05,dcb07}. The second one
is to directly detect the ionized bubbles around sources, either
through blind surveys or via targetted observations 
\cite{wlb05,aa07,gw08,mgfc07,dbc07,dmbc08}.

The major difficulty in obtaining the cosmological signal from 
these experiments is that it  is expected to be only a 
small contribution buried deep in the emission
from other astrophysical sources (foregrounds) and in the system
noise \cite{swmd99,dpar02,om03,cf04,sck05,gnb08,abc08}. It is thus a big challenge to detect the signal which is
of cosmological importance
from the other contributions that are orders of magnitude larger.
Once such challenges are dealt with, this probe will be the strongest
probe for not only reionization, but of the matter distribution
at very small scales during the dark ages.

\section{Concluding remarks}

We have discussed the
analytical approaches to model different aspects of reionization which
will help in understanding the most relevant physical processes.
In an explicit example, we have shown how to apply this formalism
for constraining the reionization history using a variety of
observational data.
These constraints imply that reionization is an extended
process over a redshift range $15 > z > 6$. It is most likely
driven by the first sources which form in small mass haloes.
However, there are still uncertainties about the exact nature of
these sources and the detailed topology of ionized regions. 
Such details are going to be addressed in near future 
as new observations, both space-borne and ground-based, 
are likely to settle these long-standing questions.
From the theoretical point of view, 
it is thereby important to develop
detailed analytical and numerical models to extract the maximum
information about the physical processes relevant for reionization 
out of the expected large and complex data sets.




\bibliography{mnrasmnemonic,reionization,astropap-mod}
\bibliographystyle{prsty}

\end{document}